\documentclass[prc,twocolumn,showpacs,preprintnumbers]{revtex4}
\usepackage{bm}
\usepackage{amssymb}
\usepackage{amsmath}
\usepackage{graphicx}

\begin{document}
\title{Thermal conductivity and phase separation of the crust of accreting neutron stars}
\author{C. J. Horowitz}\email{horowit@indiana.edu} 
\author{O. L. Caballero}\email{lcaballe@indiana.edu}
\affiliation{Department of Physics and Nuclear Theory Center,
             Indiana University, Bloomington, IN 47405}
\author{D. K. Berry}\email{dkberry@indiana.edu}
\affiliation{University Information Technology Services,
             Indiana University, Bloomington, IN 47408}

\date{\today}
\begin{abstract}
Recently, crust cooling times have been measured for neutron stars after extended outbursts.  These observations are very sensitive to the thermal conductivity $\kappa$ of the crust and strongly suggest that $\kappa$ is large.  We perform molecular dynamics simulations of the structure of the crust of an accreting neutron star using a complex composition that includes many impurities.  The composition comes from simulations of rapid proton capture nucleosynthesys followed by electron captures.  We find that the thermal conductivity is reduced by impurity scattering.  In addition, we find phase separation.  Some impurities with low atomic number $Z$ are concentrated in a subregion of the simulation volume.  For our composition, the solid crust must separate into regions of different compositions.  This could lead to an asymmetric star with a quadrupole deformation. Observations of crust cooling can constrain impurity concentrations.
\end{abstract}
\smallskip
\maketitle

\section{Introduction}
What is the thermal conductivity of the crust of a neutron star?  Recently the cooling of two neutron stars has been observed after extended outbursts \cite{Wijnands, cackett}.  These outbursts heat the stars' crusts out of equilibrium and then the cooling time is measured as the crusts return to equilibrium.  The surface temperature of the neutron star in KS 1731-260 decreased with an exponential time scale of 325 $\pm$ 100 days while MXB 1659-29 has a time scale of 505 $\pm$ 59 days \cite{cackett}.  Comparing these observations, of rapid cooling, to calculations by Rutledge et al. \cite{rutledge} and Shternin et al. \cite{shternin} strongly suggest that the crust has a high thermal conductivity.  This would be expected if the crust is a regular crystal.

In contrast, a low crust thermal conductivity, that would be expected if the crust is an amorphous solid, could help explain superburst ignition.  Superbursts are very energetic X-ray bursts from accreting neutron stars that are thought to involve the unstable thermonuclear burning of carbon \cite{superbursts, superbursts2}.  However, some simulations do not reproduce the conditions needed for carbon ignition because they have too low temperatures \cite{superignition}.  A low thermal conductivity could better insulate the outer crust and allow higher carbon ignition temperatures.

The thermal conductivity is dominated by heat conduction by electrons and this is limited by electron-ion scattering \cite{thermalcon}.  Therefore in this paper, we present molecular dynamics simulations of the crust in order to calculate electron-ion scattering.  We include many impurities based on results of a rapid proton capture nucleosynthesis simulation \cite{rpash} followed by calculations of electron capture \cite{gupta}.  We find a regular crystal structure.  We do not find an amorphous phase.    We calculate the static structure factor $S(q)$, that describes electron-ion scattering, and from $S(q)$ we determine the thermal conductivity.  Impurities can limit the thermal conductivity.  If the impurities are weakly correlated than their effect on the thermal conductivity can be described by an impurity parameter $Q$ \cite{impurities},
\begin{equation}
Q=(\Delta Z)^2=\ \langle Z^2 \rangle - \langle Z \rangle^2.
\label{Q}
\end{equation}
This depends on the dispersion in the charge $Z$ of each ion.  The rp process ash composition of ref. \cite{gupta} and ref. \cite{horowitz} has a relatively large value of $Q=38.9$.  Impurity scattering can be important at low temperatures where there is small scattering from thermal fluctuations.  Note that ref. \cite{impurities} assumes the impurities are weakly correlated.  If there are important correlations among the impurities, for example if there is a tendency for low $Z$ ions to cluster together instead of being distributed at random throughout the lattice, then the effects of impurities on the thermal conductivity could be different from what is calculated in ref. \cite{impurities}.  In this paper we perform MD simulations to study the distribution of impurities and their effect on the conductivity.

If the thermal conductivity is high, one may need additional heat sources in the crust in order to explain superburst ignition.  Although Gupta et al. \cite{gupta} find some heating from electron captures to excited nuclear states, simple nuclear structure properties may provide a natural limit to the total heating from electron captures \cite{brownprivate}.    Horowitz et al. \cite{horowitz08} find additional heating from fusion of neutron rich light nuclei such as $^{24}$O+$^{24}$O at densities near 10$^{11}$ g/cm$^3$.  These fusion reactions are an important area for future work.  Alternatively chemical separation with freezing, that was found in ref. \cite{horowitz}, could enrich the neutron star ocean with low $Z$ elements and make it easier for superburst ignitition.

In section \ref{MD} we describe our molecular dynamics simulations and the calculation of the thermal conductivity.  Results for the structure of the crust, the static structure factor $S(q)$, and the thermal conductivity are presented in section \ref{Results}.  We conclude in section \ref{Conclusions}.

\section{Molecular Dynamics Simulations}
\label{MD}
In this section we describe our molecular dynamics simulations and how we calculate the thermal conductivity.  We begin with a discussion of our initial composition.  

\subsection{Compositon}
\label{composition}

Our model for the composition of the crust is the same as was used in previous work on chemical separation when the crust freezes \cite{horowitz}.  Schatz et al. have calculated the rapid proton capture (rp) process of hydrogen burning on the surface of an accreting neutron star \cite{rpash}, see also \cite{rpash2}.  This produces a variety of nuclei up to mass $A\approx 100$.  Gupta et al. \cite{gupta} then calculate how the composition of this rp process ash evolves, because of electron capture and light particle reactions, as the material is buried by further accretion.  Their final composition, at a density of $2.16\times 10^{11}$ g/cm$^3$, has forty \% of the ions with atomic number $Z=34$, while an additional 10\% have $Z=33$.  The remaining 50\% have a range of lower $Z$ from 8 to 32.  In particular about 3\% is $^{24}$O and 1\% $^{28}$Ne.  This Gupta et al. composition \cite{gupta} is listed in Table \ref{tablezero}.  In general, nuclei at this depth in the crust are expected to be neutron rich because of electron capture.

\begin{table}
\caption{Abundance $y_z$ (by number) of chemical element $Z$.} 
\begin{tabular}{ll}
$Z$ & Abundance \\
8 & 0.0301 \\
10 & 0.0116 \\
12 & 0.0023 \\
14 & 0.0023 \\
15 & 0.0023 \\
20 & 0.0046 \\
22 & 0.0810 \\
24 & 0.0718 \\
26 & 0.1019 \\
27 & 0.0023 \\
28 & 0.0764 \\
30 & 0.0856 \\
32 & 0.0116 \\
33 & 0.1250 \\
34 & 0.3866 \\
36 & 0.0023 \\
47 & 0.0023 \\
\end{tabular} 
\label{tablezero}
\end{table}

Material accretes into a liquid ocean.  As the density increases near the bottom of the ocean, the material freezes.  However we found chemical separation when the complex rp ash mixture freezes \cite{horowitz}.  The ocean is greatly enriched in low $Z$ elements compared to the newly formed solid.  What does chemical separation mean for the structure of the crust?  Perhaps the most conservative possibility is the following steady state scenario.  We assume material accretes at a constant rate.  Initially, chemical separation enriches the ocean in low $Z$ elements.  Eventually the ocean becomes so enriched that the composition of low $Z$ material in the newly forming solid is equal to that in the accreting material.  The system reaches a steady state.  The rate of low $Z$ material accreting into the ocean is equal to the rate freezing out (modulo nuclear reactions).  The sole effect of chemical separation is to greatly enrich the ocean in low $Z$ material.  If we assume steady state, the composition of the crust will be the same as that of the original accreting material.  Therefore, in this paper we perform MD simulations to determine the structure and thermal conductivity of crust with the original Gupta et al. rp ash composition.

\subsection{Simulations of Crust Structure}
\label{simulations}

In order to calculate the thermal conductivity of a multicomponent system one needs to understand its state.  Monte Carlo simulations \cite{mcocp} of the freezing of a classical one component plasma (OCP) indicate that it can freeze into imperfect body centered cubic (bcc) or face-centered cubic (fcc) microcrystals.  Unfortunetly not much has been published on the freezing of a multi-component plasma (MCP).  There are many possibilities for the state of a cold MCP \cite{yakovlev}.  It can be a regular MCP lattice; or microcrystals; or an amorphous, uniformly mixed structure; or a lattice of one phase with random admixture of other ions; or even an ensemble of phase separated domains.  We perform classical MD simulations to explore the state of our MCP solid.  

The electrons form a very degenerate relativistic electron gas that slightly screens the interaction between ions.  We assume the potential $v_{ij}(r)$ between the ith and jth ion is,
\begin{equation}
v_{ij}(r) = \frac{Z_i Z_j e^2}{ r} {\rm e}^{-r/\lambda_e}\, ,
\label{vij}
\end{equation}
where $r$ is the distance between ions and the electron screening length is $\lambda_e=\pi^{1/2}/[2e(3\pi^2 n_e)^{1/3}]$.  Here $n_e$ is the electron density.  Note that we do not expect our results to be very sensitive to the electron screening length.  For example, the OCP melting point that we found in ref. \cite{horowitz}, using a finite $\lambda_e$, agrees well with the result for $\lambda_e=\infty$. 

To characterize our simulations , we define an average Coulomb coupling parameter $\Gamma$ for the MCP,
\begin{equation}
\Gamma= \frac{\langle Z^{5/3} \rangle \langle Z \rangle^{1/3} e^2}{a T}\, ,
\label{gamma}
\end{equation}
where the mean ion sphere radius is $a=(3/4\pi n)^{1/3}$ and $n=n_e/\langle Z\rangle$ is the ion density.   The OCP freezes at $\Gamma=175$.  In ref. \cite{horowitz} we found that the impurities in our MCP lowered the melting temperature until $\Gamma=247$.  Finally, we can measure time in our simulation in units of one over an average plasma frequency $\omega_p$,
\begin{equation}
\omega_p=\Bigl(\sum_j\frac{Z_j^2 4\pi e^2 x_j n}{M_j}\Bigr)^{1/2}\, ,
\label{omegap}
\end{equation}
where $M_j$ is the average mass of ions with charge $Z_j$ and abundance $x_j$ (by number).  Note that there will be quantum corrections to our classical simulations for temperatures significantly below the plasma frequency.

\subsection{Thermal conductivity}
\label{Thermal conductivity}
The thermal conductivity $\kappa$ has been discussed by Potekhin et al. \cite{thermalcon}.  We assume $\kappa$ is dominated by heat carried by electrons \cite{thermalcon},
\begin{equation}
\kappa=\frac{\pi^2 k_B^2 T n_e}{3 m_e^*\, \nu},
\label{kappa}
\end{equation}
where the effective electron mass is $m_e^*=\epsilon_F=(k_F^2+m_e)^{1/2}$ with $k_F$ the electron Fermi momentum and $m_e$ the electron mass.  The electron collision frequency $\nu$ is assumed to be dominated by electron-ion collisions \cite{thermalcon},
\begin{equation}
\nu=\frac{4}{3\pi}\langle Z \rangle \epsilon_F \alpha^2 \Lambda\, .
\end{equation}
Here $\alpha$ is the fine structure constant and $\Lambda$ is the Coulomb logarithm that describes electron-ion collisions \cite{thermalcon},
\begin{equation}
\Lambda=\int_{q_0}^{2k_F} \frac{dq}{q \epsilon(q,0)^2} S'(q) (1-\frac{q^2}{4k_F^2})\, .
\label{Lambda}
\end{equation}
Here $\epsilon(q,0)$ is the dielectric function due to degenerate relativistic electrons, \cite{jancovici}, see Eq. 2.3 of \cite{itoh}.  Note that for simplicity we neglect second and higher Born corrections to electron ion scattering in Eq. \ref{Lambda}, see for example \cite{itoh}.  We are interested in the difference in thermal conductivity for different solid structures.   Second and higher Born corrections should be the same for the different structures.  Finally, the lower limit $q_0$ in Eq. \ref{Lambda} is $0$ in a liquid phase and $q_0=(6\pi^2n)^{1/2}$ in a crystal phase \cite{thermalcon}.  

The static structure factor $S(q)$ describes electron-ion scattering.  We calculate $S(q)$ directly as a density-density correlation function using trajectories from our MD simulations,
\begin{equation}
S({\bf q})=\langle \rho^*({\bf q})\rho({\bf q}) \rangle - |\langle \rho({\bf q})\rangle |^2\, .
\label{S(q)}
\end{equation}
Here the charge density $\rho({\bf q})$ is,
\begin{equation}
\rho({\bf q}) = \frac{1}{\sqrt N} \sum_{i=1}^N \frac{Z_i}{\langle Z \rangle} {\rm e}^{i {\bf q \cdot r}_i},
\label{rho}
\end{equation}
with $N$ the number of ions in the simulation and $Z_i$, ${\bf r}_i$ are the charge and location of the ith ion.  We evaluate the thermal average in Eq. \ref{S(q)} as a time average during our MD simulations.

The static structure factor $S(q)$ includes both Bragg scattering contributions from the whole crystal lattice $S_{\rm bragg}(q)$ and inelastic excitation contributions  $S'(q)$ \cite{thermalcon},
\begin{equation}
S({\bf q}) = S'({\bf q}) + S_{\rm bragg}({\bf q})\, .
\label{S'(q)}
\end{equation}
The Bragg contribution is a series of delta functions at momenta related to one over the lattice spacing.  This describes Bragg scattering and helps determine the electron band structure.  It does not limit the electron mean free path.  Instead the mean free path and thermal conductivity are determined by $S'({\bf q})$.

Our MD simulations are classical.  Unfortunately this classical approximation makes the separation of $S(q)$ into $S'$ and $S_{\rm bragg}$ somewhat ambiguous.  We approximate $S'({\bf q})$ with a simple numerical filter applied to $S({\bf q})$.  The filter removes delta function like contributions to $S({\bf q})$ that have a very rapid $q$ dependence, and also removes numerical noise.  This is discussed further in Section \ref{Results}. 
 
\section{Results}
\label{Results}

To explore possible states for the multicomponent plasma we perform two molecular dynamics simulations.  The initial conditions of these simulations are similar to those in \cite{horowitz}.  The composition is indicated in Table \ref{tablezero}.  We start by freezing a very small system of 432 ions.  Here the ions were started with random initial conditions at a high temperature $T$ and $T$ was reduced in stages (by re-scaling velocities) until the system freezes.  For the first simulation run, called rpcrust-01b in Table \ref{tabletwo}, we place four copies of this 432 ion solid in a larger simulation volume along with four copies of a 432 ion liquid configuration.  This 3456 ion configuration is evolved at a lower temperature until the whole system freezes.  Next, we evolve the 3456 ion solid at a reference density of $n=7.18\times 10^{-5}$ fm$^{-3}$ (or $1\times 10^{13}$ g/cm$^3$) for a total simulation time of $2.4\times 10^9$ fm/c ($8.9\times 10^6$ $\omega_p^{-1}$).   The temperature was started at 0.325 MeV and slowly decreased to a small value by the end of this time.  The density and initial temperature correspond to $\Gamma=261.6$.  Evolution was done using the velocity verlet algorithm \cite{verlet} using a time step of $\Delta t=25$ fm/c for a total of $9.6\times 10^7$ steps.  This took about 2 months on a single special purpose MDGRAPE-2 \cite{mdgrape} board.  
Next, this low temperature configuration was reheated to $T=0.325$ MeV and evolved for $1.6\times 10^9$ fm/c.  The total time was $4\times 10^9$ fm/c.  This somewhat complicated procedure was done for historical reasons.  It does allow plenty of time for ions to diffuse throughout the simulation volume.

Note that at our artificially high reference density ($10^{13}$ g/cm$^3$) free neutrons will be present.  However, we are primarily interested in lower densities with out free neutrons.  Our results can be scaled to other densities and temperatures such that the Coulomb parameter $\Gamma$ remains the same, see below.  Furthermore, although we quote all simulation times in fm/c, the times can be expressed in terms of one over the average plasma frequency using $1/\omega_p=270$ fm/c.

\begin{table}
\caption{Computer Simulations.  The start time is $t_i$, the finish time $t_f$, $N$ is the number of ions, and the temperature is $T$.  Each simulation is at a density $n=7.18\times 10^{-5}$ fm$^{-3}$ ($1\times 10^{13}$ g/cm$^3$).  Note that one over the plasma frequency is $1/\omega_p=270$ fm/c.} 
\begin{tabular}{lllll}
Run & $N$ & $t_i$(fm/c)& $t_f$(fm/c) & $T$(MeV) \\
\toprule
rpcrust-01b &3456 &  0& $1.6\times 10^9$& 0.325  \\
rpcrust-05 & 3456  & 0 & $4\times 10^8$ & 0.1 \\
 & & $4 \times 10^8$ & $ 8\times 10^8$ & 0.2 \\
  & & $8\times 10^8$ & $1.2\times 10^9$ & 0.3 \\
  OCP & 1024 & 0 & $1.6\times 10^7$ & 0.334  \\
\end{tabular} 
\label{tabletwo}
\end{table}

\begin{figure}[ht]
\begin{center}
\includegraphics[width=3in,angle=0,clip=true] {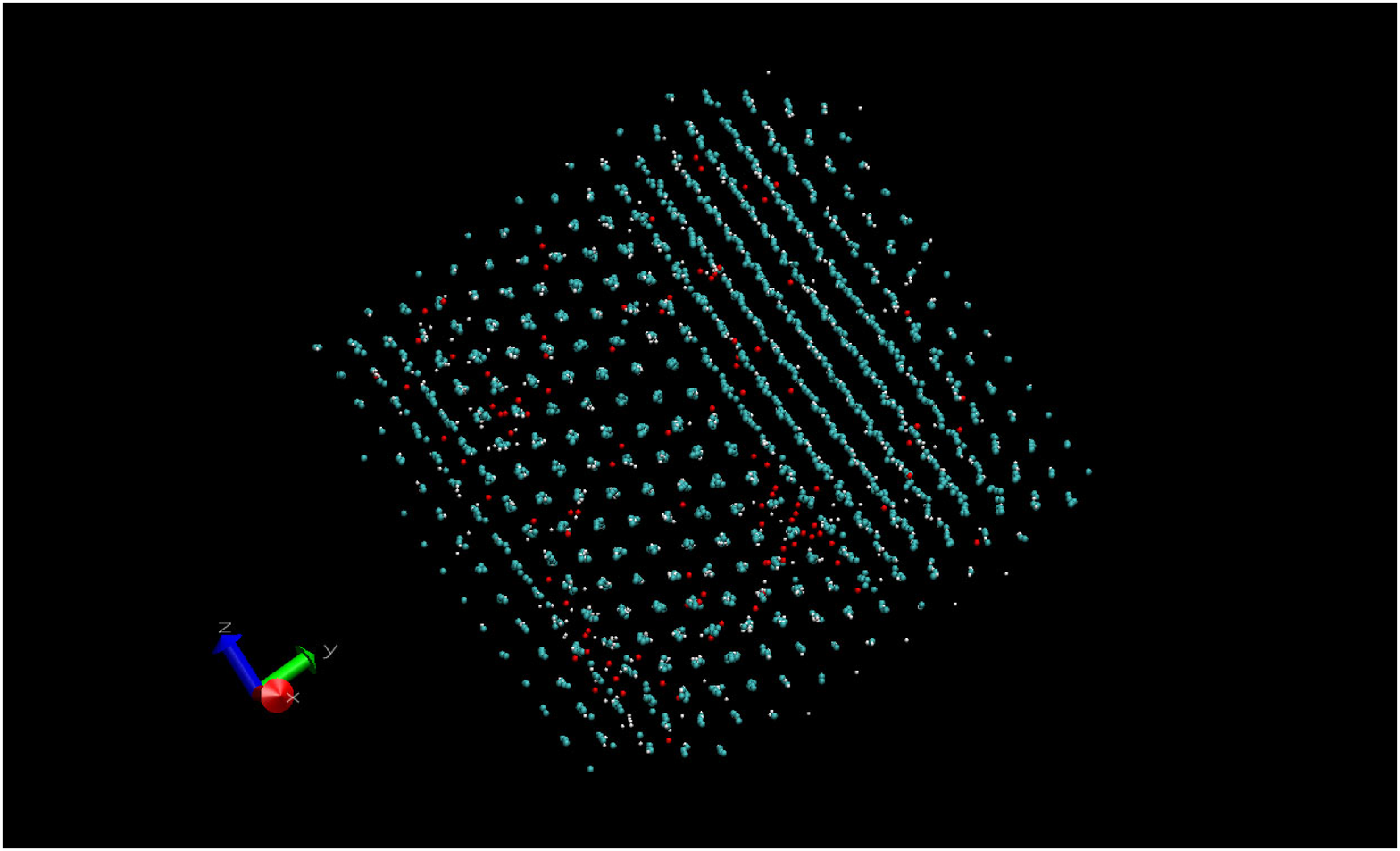}
\caption{(Color on line) Configuration of the 3456 ion mixture in run rpcrust-01b  at the start of the simulation.  The small red spheres show the positions of $^{24}$O ions, while ions of above average $Z$ are shown as larger blue spheres.  Finally, ions of below average $Z$ (except for O) are shown as small white spheres. The left and right halves of the figure show two micro-crystals of different orientations.}
\label{Fig1}
\end{center}
\end{figure}

The initial configuration for run rpcrust-01b, see Table \ref{tabletwo}, is shown in Fig. \ref{Fig1}.  The system is seen to be composed of two micro-crystals of different orientations.  This is similar to the micro-crystals found in ref. \cite{mcocp} upon freezing a one component plasma.  In Fig. \ref{Fig1} we highlight the positions of the $^{24}$O ions (as small red spheres).  These ions are located both in the crystal planes and in between them.  The O ions are not spread uniformly throughout the volume but there is a tendency for them to cluster.  This will be discussed in more detail below.

\begin{figure}[ht]
\begin{center}
\includegraphics[width=3in,angle=0,clip=true] {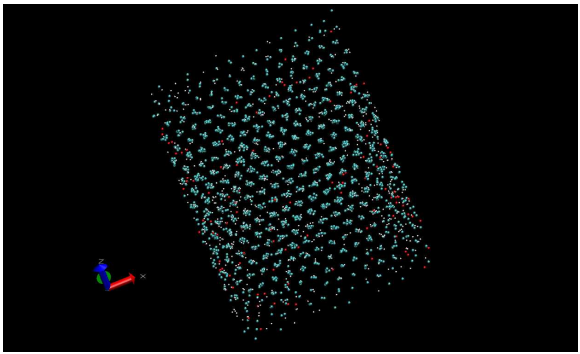}
\caption{(Color on line) Configuration of the 3456 ion mixture in run rpcrust-01b after a simulation time of $1.6 \times 10^9$ fm/c.  The small red spheres show the positions of $^{24}$O ions, while ions of above average $Z$ are shown as larger blue spheres.  Finally, ions of below average $Z$ (except for O) are shown as small white spheres.}
\label{Fig2}
\end{center}
\end{figure}

This configuration was then reheated to $T=0.325$ MeV and evolved for $1.6\times 10^9$ fm/c. The final configuration of run rpcrust-01b is shown in Fig. \ref{Fig2}.  The two micro-crystals of different orientation are now gone.  The system has managed to anneal into a single crystal with a single orientation.  This suggests that micro-crystals could be an artifact of computer simulations of limited size and duration.  It also suggests that neutron star crust could be formed with relatively large domain sizes.  

Figure \ref{Fig2} shows that O ions and other low $Z$ impurities are enhanced in regions on the left and right of the simulation volume.  Because of the periodic boundary conditions this actually corresponds to a single region.  We conclude that this complex mixture does not form a single uniform solid phase.  Instead it separates into two solid phases.  One phase is enriched in high $Z$ ions and the other phase is enriched in low $Z$ ions.

To study this further we have performed another simulation labeled rpcrust-05 in Table \ref{tabletwo}.  The starting point was similar to run rpcrust-01b with eight copies of a 432 ion configuration placed into a larger simulation volume.  This 3456 ion configuration was evolved for $2.5\times 10^9$ fm/c as the temperature was slowly decreased from $0.35$ MeV to a small value.  Next, this low temperature configuration was heated to $T=0.1$ MeV and evolved for 400 million fm/c, the system was then heated to $T=0.2$ MeV and evolved for another 400 million fm/c and finally the system was heated to $T=0.3$ MeV and evolved for a final 400 million fm/c as indicated in Table \ref{tabletwo}.   The total simulation time including both the original preparation and the $T=0.1$, 0.2, and 0.3 MeV runs was $3.7\times 10^9$ fm/c.  

\begin{figure}[ht]
\begin{center}
\includegraphics[width=3in,angle=0,clip=true] {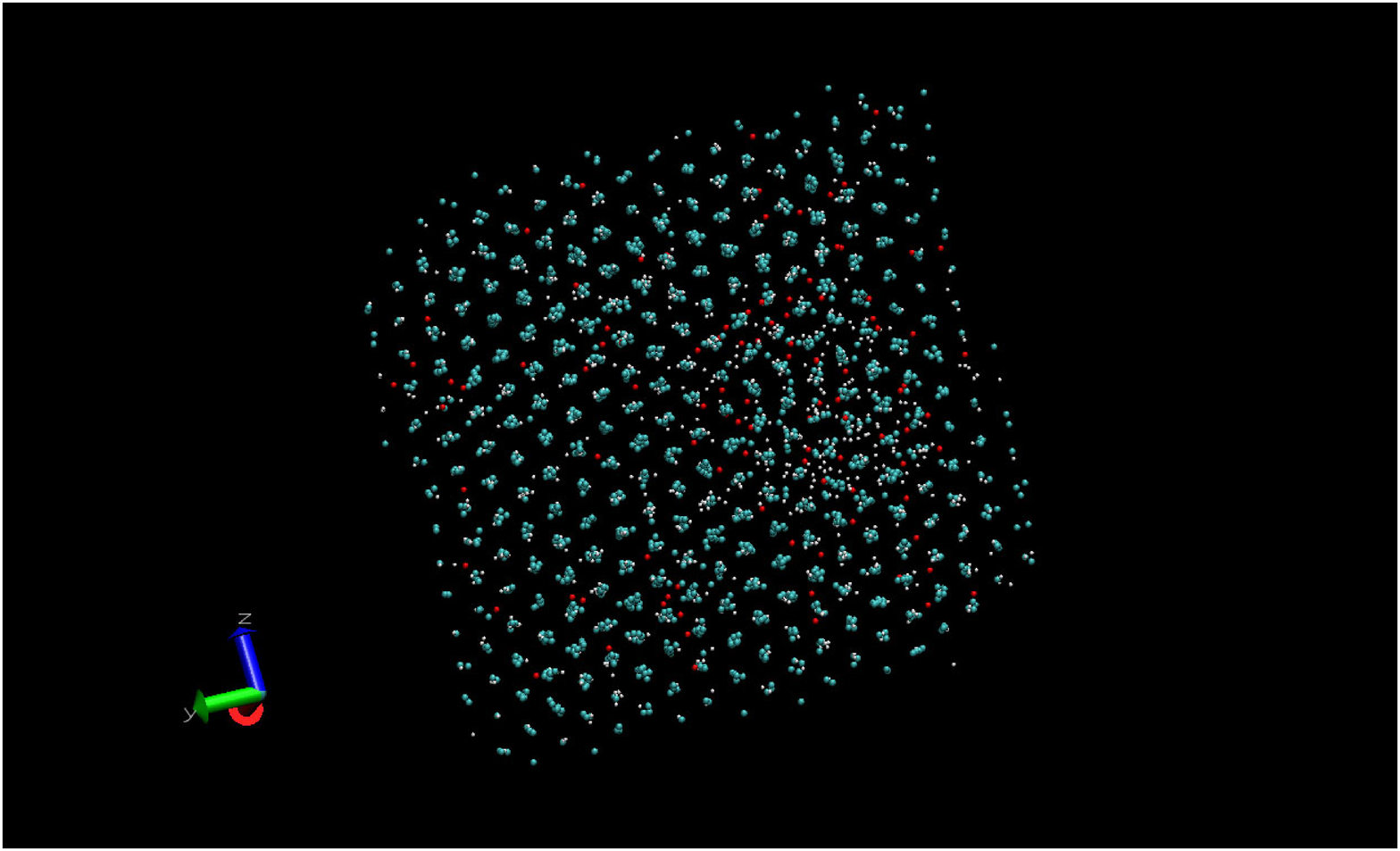}
\caption{(Color on line) Configuration of the 3456 ion mixture in run rpcrust-05 after a simulation time of $1.2 \times 10^9$ fm/c.  The small red spheres show the positions of $^{24}$O ions, while ions of above average $Z$ are shown as larger blue spheres.  Finally, ions of below average $Z$ (except for O) are shown as small white spheres.  The $^{24}$O concentration is seen to be enhanced in a sub-region to the right of center.}
\label{Fig3}
\end{center}
\end{figure}

\begin{figure}[ht]
\begin{center}
\includegraphics[width=2.75in,angle=270,clip=true] {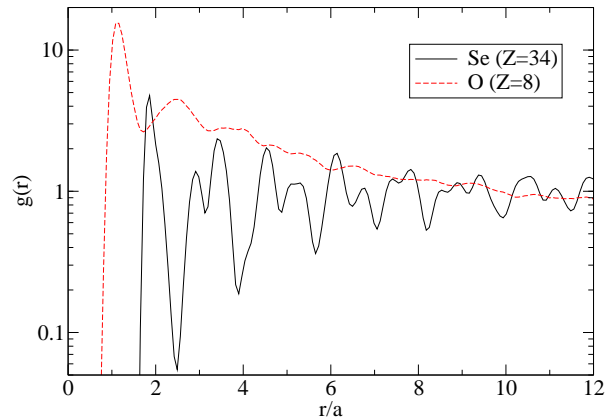}
\caption{(Color on line) Radial distribution function $g(r)$ versus $r$ over the mean ion sphere radius $a$, for run rpcrust-05 at a temperature $T=0.1$ MeV.  The dashed (red) line shows the correlation function between O ions while the solid (black) line shows the Se-Se correlation function. }
\label{Fig3b}
\end{center}
\end{figure}

The final configuration of run rpcrust-5 is shown in Fig. \ref{Fig3}.  The system involves only a single body-centered cubic (bcc) crystal.  However O and other low $Z$ ions are not uniformly distributed.  Instead they are strongly enriched in a local region.  This is indicated in Fig. \ref{Fig3b} that shows the radial distribution function $g(r)$ for run rpcrust-05 at a temperature $T=0.1$ MeV.  The peaks in the Se-Se correlation function show the regular lattice planes.  However $g(r)$ for O-O is seen to be larger than one over a range of moderate distances $r$.  This shows that the O ions are concentrated in a localized sub-volume.  We conclude that the complex rp ash mixture does not form a single solid phase.  Instead, for this composition,  the neutron star crust must be composed of two or more regions of different compositions.  This disproves our steady state assumption.  There appears to be no composition of the liquid ocean, no matter how enriched in low $Z$ ions, that allows a uniform solid phase to form.  

These multiple regions of the crust with different compositions may be very important for the structure of the neutron star.  For example, if the phases are not distributed uniformly, this could lead to a mass quadruple moment that might radiate gravitational waves \cite{crustmonster}.   This nonuniform distribution of phases could arise from an anisotropic temperature because phase separation is temperature dependent.

Finally for comparison we have also performed a one component plasma simulation, see run OCP in Table \ref{tabletwo},  where each ion has a charge $Z=29.4$ equal to the average charge in the MCP simulations.  Simulation OCP started from a random configuration of 1024 ions and the temperature was reduced in stages until $\Gamma=300$ at which point the simulation was observed to freeze.  Finally this solid was heated up to $\Gamma=250$ for the final results.

\subsection{Static Structure Factor}

We calculate the static structure factor $S(q)$ from the density-density correlation function, Eq. \ref{S(q)}.  The thermodynamic average is approximated as a time average over $6.25\times 10^6$ fm/c of simulation time.  We present results for the angle averaged $S(q)$ after averaging over approximately 50 different directions of $\vec q$.  These results are somewhat time consuming because we calculate $S(q)$ for approximately 1400 different values of $|\vec q|$ for run rpcrust05.

We calculate the inelastic contribution $S'(q)$ by applying a simple numerical filter that removes very rapid changes in $S(q)$ with $q$.  Our filter, applied to a table of $q_i$ and $S(q_i)$ values, works as follows:  if $S(q_i)$ differs by more than some threshold $\approx 0.1$ from $S(q_{i-1})$ than $q_i$ and $S(q_i)$ are removed from the table.  This removes numerical noise and may remove delta function like contributions from the Bragg peaks.    In addition, we may simply miss some Bragg peaks because we only calculate $S(q)$ for a finite number of $q$ points.  Our motivation for this simple procedure is to calculate $S'(q)$ and $\Lambda$ based on $S(q)$ calculations that are not likely contaminated by Bragg contributions.  

\begin{figure}[ht]
\begin{center}
\includegraphics[width=2.8in,angle=270,clip=true] {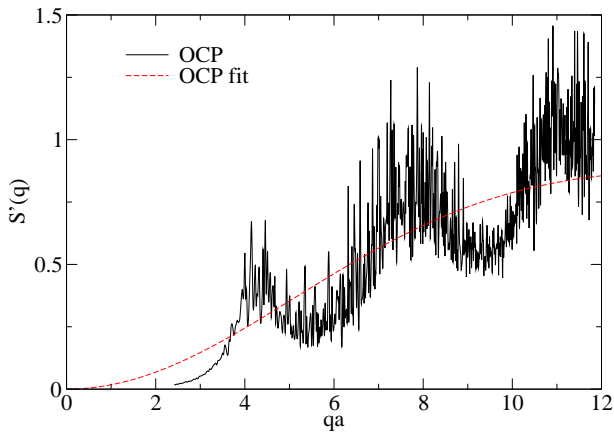}
\caption{(Color on line) Inelastic contributions to the static structure factor $S'(q)$ versus momentum transfer $q$ times the mean ion sphere radius $a$, for run OCP, solid black line and the simple fit presented in ref. \cite{thermalcon} , dashed red line.}
\label{Fig4}
\end{center}
\end{figure}

We first test this procedure with the one component plasma simulation OCP of Table \ref{tabletwo}, see Fig. \ref{Fig4}.  Our results for $S'(q)$ show more structure than the simple fit presented in ref. \cite{thermalcon}.   Note that this may reflect a limitation of the fit.  In addition, there is some high frequency noise in our simulation.  However, there is good agreement, to  4 \%, between our OCP simulation and the fit for the integral of $S'(q)$ over $q$ that is needed to calculate the Coulomb logarithm $\Lambda$, see Eq. \ref{Lambda} and Table \ref{tablethree}.  Therefore, our procedure for $S'(q)$ reproduces the known Coulomb logarithm and thermal conductivity of a one component plasma.  

\begin{table}
\caption{Coulomb Logarithm, Eq. \ref{Lambda}, for a one component plasma (OCP).  The $\Lambda_{\rm OCPfit}$ value is from the $S'(q)$ fit in ref. \cite{thermalcon} while $\Lambda_{\rm OCP}$ is our calculation for simulation OCP. } 
\begin{tabular}{lll}
$\Gamma$ & $\Lambda_{\rm OCPfit}$ & $\Lambda_{\rm OCP}$\\
\toprule
250 & 0.362 & 0.348 \\
\end{tabular} 
\label{tablethree}
\end{table}

\begin{figure}[ht]
\begin{center}
\includegraphics[width=2.8in,angle=270,clip=true] {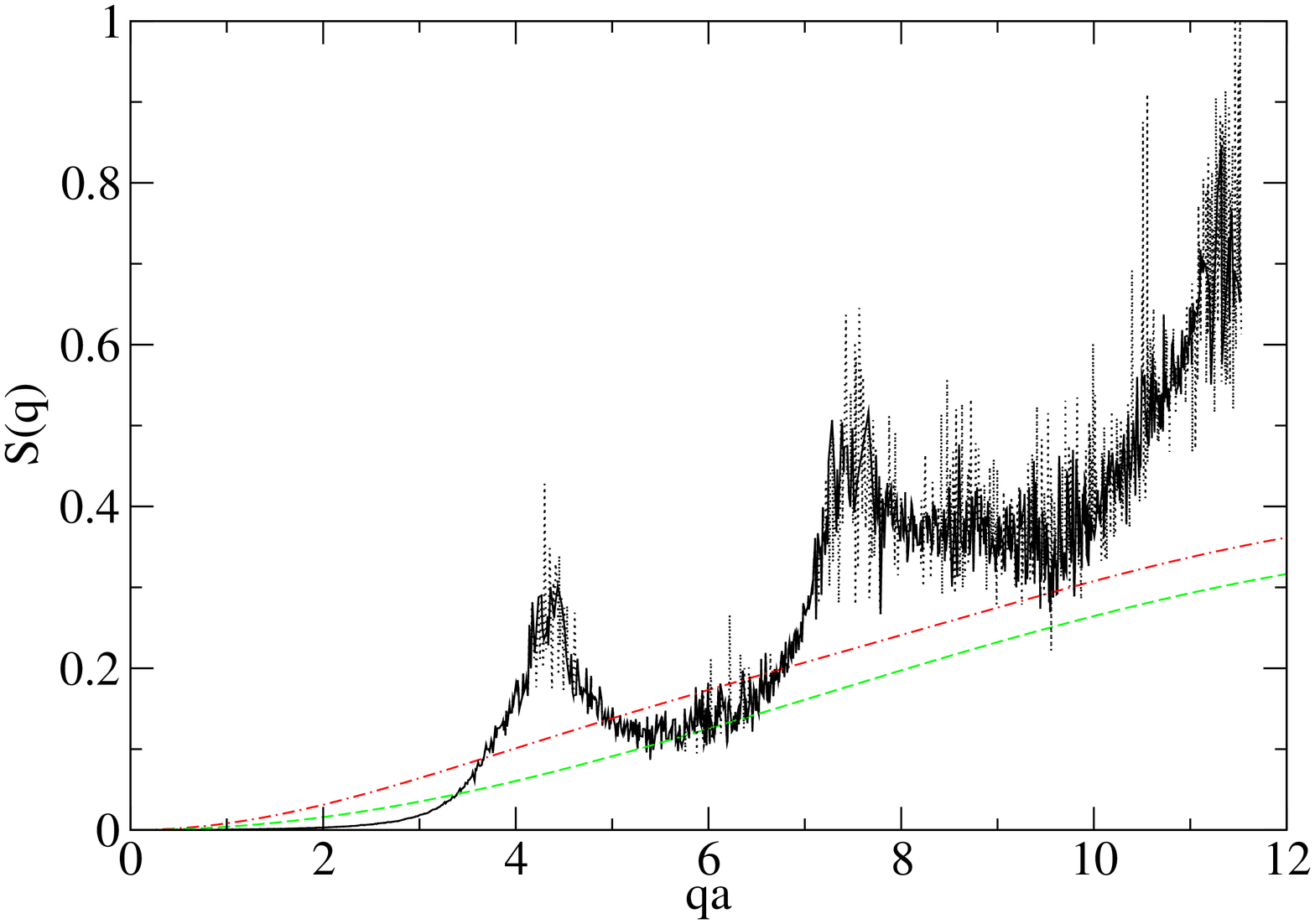}
\caption{(Color on line) Static structure factor $S(q)$ for run rpcrust-05 versus momentum transfer $q$ times the mean ion sphere radius $a$, dotted black line, at a temperature $T=0.1$ MeV.  The solid black line is an approximation to the inelastic contribution $S'(q)$.  This is calculated with a simple numerical filter applied to $S(q)$.  Finally the  dashed green line is the fit to OCP results for $S'(q)$ from ref. \cite{thermalcon} and the dashed dotted red line adds impurity scattering from ref. \cite{impurities} to these OCP results.  }
\label{Fig5}
\end{center}
\end{figure}

\begin{figure}[ht]
\begin{center}
\includegraphics[width=2.8in,angle=270,clip=true] {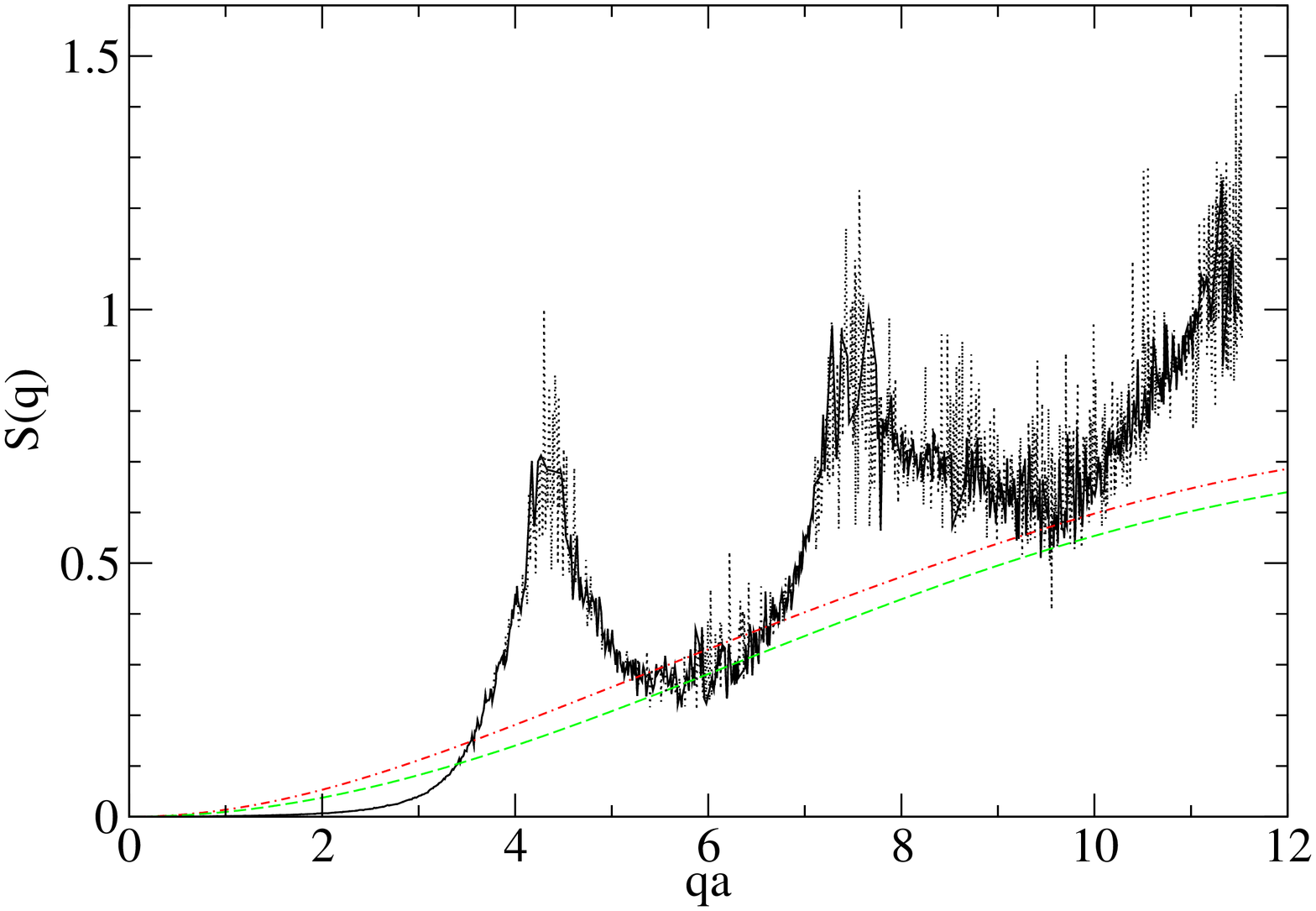}
\caption{(Color on line) Static structure factor $S(q)$ for run rpcrust-05 versus momentum transfer $q$ times the mean ion sphere radius $a$, dotted black line, at a temperature $T=0.2$ MeV.  The solid black line is an approximation to the inelastic contribution $S'(q)$.  This is calculated with a simple numerical filter applied to $S(q)$.  Finally the  dashed green line is the fit to OCP results for $S'(q)$ from ref. \cite{thermalcon} and the dashed dotted red line adds impurity scattering from ref. \cite{impurities} to these OCP results.  }
\label{Fig6}
\end{center}
\end{figure}

\begin{figure}[ht]
\begin{center}
\includegraphics[width=2.8in,angle=270,clip=true] {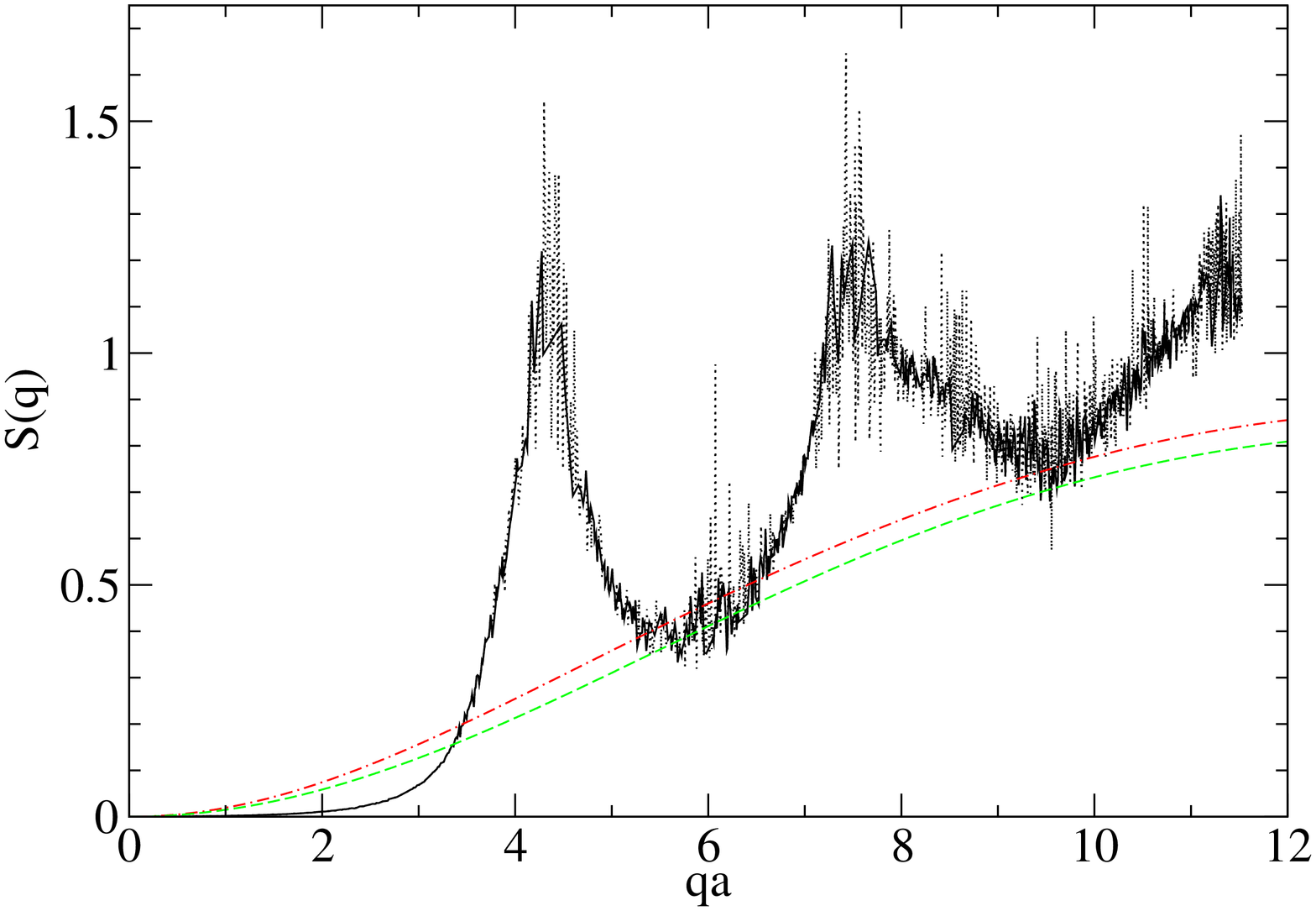}
\caption{(Color on line) Static structure factor $S(q)$ versus momentum transfer $q$ times the mean ion sphere radius $a$ for run rpcrust-05, dotted black line, at a temperature $T=0.3$ MeV.  The solid black line is an approximation to the inelastic contribution $S'(q)$.  This is calculated with a simple numerical filter applied to $S(q)$.  Finally the  dashed green line is the fit to OCP results for $S'(q)$ from ref. \cite{thermalcon} and the dashed dotted red line adds impurity scattering from ref. \cite{impurities} to these OCP results.  }
\label{Fig7}
\end{center}
\end{figure}

Figures \ref{Fig5}, \ref{Fig6}, and \ref{Fig7}  show $S(q)$ for run rpcrust-05 at temperatures of $T=0.1$, 0.2, and 0.3 MeV respectively.   We expect similar results for run rpcrust-01b.  These figures also show  the simple fit to $S'(q)$ for an OCP presented in ref. \cite{thermalcon}.  This fit is significantly below $S'(q)$ for run rpcrust-05.  Finally, these figures show the contribution of impurity scattering from \cite{impurities} added to the OCP fit results.  Impurity scattering depends on $Q$, see Eq. \ref{Q} and $Q=38.9$ for run rpcrust-05.  We find that results for run rpcrust-05 are still above the OCP fit even when impurity scattering is added.  Note, that impurity scattering is automatically included in our MD simulation because of the complex composition used.  Table \ref{tablefour} presents Coulomb logarithms $\Lambda$ for rp ash composition.  Again, results for run rpcrust-05 are above the OCP plus impurities calculation.  However the difference is only 18\% at a temperature of 0.1 MeV.  Note in Reference \cite{impurities} it was explicitly assumed that the impurities are randomly distributed.  However, we find strong correlations among the impurities, see Fig. \ref{Fig3b} for example.  Therefore it is perhaps not surprising that we find larger effects from impurities than ref. \cite{impurities}. 

\begin{table}
\caption{Coulomb Logarithm, Eq. \ref{Lambda}, for rp process ash composition.  The $\Lambda$ values are from run rpcrust05 at the indicated temperatures $T$, while $\Lambda_{\rm OCPfit}$ is from ref. \cite{thermalcon} for a pure OCP and $\Lambda_{\rm OCP+imp}$ also includes impurity scattering from ref. \cite{impurities} with $Q=38.9$. } 
\begin{tabular}{lllll}
$T$(MeV) & $\Gamma$ & $\Lambda_{\rm OCPfit}$ & $\Lambda_{\rm OCP+imp}$& $\Lambda$\\
\toprule
0.1 & 850 & 0.104 & 0.146 & 0.173  \\
0.2 & 425 & 0.232 & 0.276 & 0.366 \\
0.3 & 283 & 0.334 & 0.377 & 0.530 \\
\end{tabular} 
\label{tablefour}
\end{table}

\begin{table}[htbp]
\caption{Thermal conductivity for a temperature of $T=0.043$ MeV ($5\times 10^8$K). Results have been scaled to the indicated densities.  The thermal conductivity $\kappa$ is from run rpcrust-05 while $\kappa_{\rm OCP+imp}$ is based on the OCP plus impurity values $\Lambda_{\rm OCP+imp}$ in Table \ref{tablefour} and $\kappa_{\rm OCP}$ is for the fit to an OCP without any impurity scattering. } 
\begin{tabular}{lllll}
$\Gamma$ &  $\rho$   & $\kappa_{\rm OCP}$& $\kappa_{\rm OCP+imp}$ & $\kappa$ \\
  & (g/cm$^{3}$) & (erg/K cm s) & (erg/K cm s) & (erg/K cm s) \\
\toprule
850 &  $7.91\times 10^{11}$ & $2.48\times 10^{19}$&$1.77\times 10^{19}$ & $1.49\times 10^{19}$\\
425 &  $9.89\times 10^{10}$ & $5.58\times 10^{18}$&$4.69\times 10^{18}$ & $3.54\times 10^{18}$\\
283 &  $2.92\times 10^{10}$ & $2.58\times 10^{18}$& $2.29\times 10^{18}$ & $1.63\times 10^{18}$\\
\end{tabular} 
\label{tablefive}
\end{table}

Our results for $S(q)$ in Figs. \ref{Fig4}-\ref{Fig7} show statistical noise.  However some of the effects of this noise average to zero when one integrates over $S'(q)$ to calculate $\Lambda$.  We estimate the statistical error in our calculation of $\Lambda$ at $T=0.1$ MeV, see Table \ref{tablefour}, to be $\pm$ 0.001 by comparing calculations of $\Lambda$ using configurations for simulation times of $3\times 10^8$ fm/c to $4\times 10^8$ fm/c to a calculation using configurations from $2\times 10^8$ to $3\times 10^8$ fm/c.  We emphasize that our procedure to calculate $S'(q)$ from $S(q)$ is model dependent.  Our numerical filter not only removes Bragg peaks but it may also remove some statistical noise.  Note that removing some noise seems to have minimal effects on the values of $\Lambda$ that we calculate.  We do not believe our results in Table \ref{tablefour} are very sensitive to our procedure to determine $S'(q)$.  This is based on explicit calculations with a few different procedures.

\subsection{Thermal Conductivity}

We now calculate the thermal conductivity $\kappa$ using our results for the Coulomb logarithms.  These results can be scaled to a range of densities $n$ and temperatures $T$ so that the value of $\Gamma$, Eq. \ref{gamma}, remains the same.  Table \ref{tablefive} presents $\kappa$ at a temperature of $T=5\times 10^8$ K (a typical value for a super bursting star).  The thermal conductivity is lower for run rpcrust05 than for an OCP.  First, this is because run rpcrust05 has a large number of impurities, corresponding to the large impurity parameter $Q=38.9$.  Second, we think $\kappa$ may be further reduced because the impurities in run rpcrust05 are not distributed uniformly.  Instead they are concentrated in one region.   

Although our simulations show some of the effects of impurities on the thermal conductivity, we emphasize that there may be important finite size effects because we find clustering.  It is unrealistic to describe a large system by simply repeating our small simulation volume many times.  This would describe the impurities as being concentrated into many very small regions.  Instead, we believe the concentration of impurities indicates phase separation.  We think that a large sample will separate into two (or more) bulk phases.  It is important to study phase separation further with larger molecular dynamics simulations and this may change our thermal conductivity results.  In general, one phase will be enriched in high $Z$ ions while the other is enriched in low $Z$ ions.  Phase separation may act to reduce the impurity parameter $Q$ and increase the thermal conductivity.  For example, $Q$ will be reduced in the high $Z$ phase because low $Z$ impurities have gone into the other phase.  

In addition, nuclear reactions may reduce $Q$ further.  In general, we expect nuclear reactions to preferentially burn low $Z$ impurities because of their low Coulomb barriers.  See for example ref. \cite{horowitz08}.  This will reduce $Q$ and increase the thermal conductivity.  One should study how $Q$ evolves with depth because of reactions.   Finally, it is important to analyze observations of crust cooling after extended outbursts \cite{rutledge,shternin} to see what observational constraints can be placed on the thermal conductivity and $Q$.  It may be that observations of rapid crust cooling can strongly limit the size of $Q$ \cite{andrew}.

Phase separation may have another important effect.  It will create layers in the crust of different compositions and densities.  These layers may not be spherically symmetric.  For example, phase separation depends on temperature.  Therefore an anisotropic temperature distribution will lead to an anisotropic density.  It is important to study how phase separation will change the structure of the star.

\section{Summary and Conclusions}
\label{Conclusions}
The crust of an accreting neutron star, likely, has a complex composition with many impurities.  Nuclei are synthesized via the rapid proton capture process and the composition is modified by electron capture as material is buried to greater densities.  We have performed MD simulations, with a complex composition, to study the structure of the crust.  Our simulations form ordered crystals rather than an amorphous solid.  

However, we find phase separation.  Some low $Z$ impurities are concentrated into a subregion of the full simulation volume.  This phase separation, between two solid phases, is similar to the chemical separation found previously between liquid and solid phases \cite{horowitz}.  Previously, we assumed a steady state equilibrium where chemical separation greatly increases the concentration of low $Z$ impurities in the liquid ocean.  However, the composition of the solid crust was assumed to be the same as that of the accreting material.  Our new results disprove this steady state assumption.

The crust can not be uniform, given our initial composition.  Phase separation will divide the crust into two or more regions of different compositions.  This may have important implications for the structure of the star.  For example, composition anisotropies could lead to  gravitational wave radiation from a quadrupole deformation \cite{crustmonster}.   In future work we will study the size of possible compositional asymmetries because of an anisotropic temperature distribution
     
We calculated the static structure factor $S(q)$ for our simulations and from $S(q)$ the thermal conductivity $\kappa$.  Since our simulations have a complex composition, we automatically include the contributions of impurity scattering.  We find that $\kappa$ is somewhat reduced because of impurity scattering and because the impurities are not distributed uniformly.   We expect the same results for the electrical conductivity $\sigma$, that is important for magnetic field decay \cite{Bdecay}, and the shear viscosity $\eta$, that can damp neutron star oscillations \cite{viscosity},\cite{viscospasta}.  The reduction in $\kappa$ may be observable in crust cooling times and these observations may set limits on impurities.  Future work should study how phase separation and or nuclear reactions impact impurity concentrations. 
 
\section{Acknowledgments}
We thank Ed Brown and  Andrew Cumming for helpful discussions and acknowledge the hospitality of the Institute for Nuclear Theory where this work was started.  This work was supported in part by DOE grant DE-FG02-87ER40365 and by Shared University Research grants from IBM, Inc. to Indiana University.

\vfill\eject


\begin{thebibliography}{99} 
\bibitem{Wijnands} R.~Wijnands et al., astro-ph/0405089.
\bibitem{cackett} E. M. Cackett et al., MNRAS {\bf 372}, 479 (2006). \bibitem{rutledge} R. E. Rutledge et al., ApJ. {\bf 580}, 413 (2002).
\bibitem{shternin} P. S. Shternin, D. G. Yakovlev, P. Haensel, and A. Y. Potekhin, MNRAS {\bf 382}, L43 (2007).
\bibitem{superbursts} A. Cumming and L. Bildsten, ApJ {\bf 559} (2001) L127.  
\bibitem{superbursts2} T. E. Strohmayer and E. F. Brown, ApJ {\bf 566} (2002) 1045.
\bibitem{superignition}  A. Cumming, J. Macbeth, J. J. M. in 't Zand and D. Page, ApJ. {\bf 646} (2006) 429.
\bibitem{thermalcon} A. Y. Potekhin, D. A. Baiko, P. Haensel, and D. G. Yakovlev, Astron.  Astrophys., {\bf 346} (1999) 345.
\bibitem{rpash} H. Schatz et al., PRL {\bf 86} (2001) 3471.
\bibitem{gupta} S. Gupta, E. F. Brown, H. Schatz, P. Moller, and K-L. Kratz, ApJ {\bf 662} (2007) 1188.
\bibitem{impurities} N. Itoh and Y. Kohyama, ApJ. {\bf 404} (1993) 268.
\bibitem{horowitz} C. J. Horowitz, D. K. Berry, and E. F. Brown, PRE {\bf 75} (2007) 066101.
\bibitem{brownprivate} E. F. Brown, private communication.
\bibitem{horowitz08} C. J. Horowitz, H. Dussan and D. K. Berry, PRC {\bf 77} (2008) 045807. 
\bibitem{rpash2} S. E. Woosley, A. Hager, A. Cumming, R. D. Hoffman, J. Pruet, T. Rauscher, J. L. Fisker, H. Schatz, B. A. Brown, and M. Wiescher, ApJ Supp. {\bf 151} (2004) 75.
\bibitem{mcocp}H. E. Dewitt, W. L. Slattery, and J. Yang in ``Strongly Coupled Plasmas'', eds. H. M. Van Horn and S. Ichimaru, Univ. of Rochester Press 1993, p425.
\bibitem{yakovlev} D. G. Yakovlev, L. R. Gasques, M. Wiescher, and A. V. Afanasjev, PRC {\bf 74} (2006) 035803.
\bibitem{jancovici} B. Jancovici, J. Stat. Phys. {\bf 17} (1977) 357.
\bibitem{itoh} N. Itoh, S. Uchida, Y. Sakamoto, and Y. Kohyama, arxiv:0708.2967.
\bibitem{verlet} L. Verlet, Phys. Rev. {\bf 159}, 98 (1967).
F. Ercolessi, A Molecular Dynamics Primer, available from http://www.sissa.it/furio/ (1997).
\bibitem{mdgrape} J. Makino, T. Fukushige, M. Koga, and E. Koutsofias, in Proceeding of SC2000, Dallas, 2000.
\bibitem{andrew}A. Cumming and E. F. Brown, private communication.
\bibitem{crustmonster} G. Ushomirsky, C. Cutler, and L. Bildsten, MNRAS {\bf 319} (2000) 902.
\bibitem{viscosity} A. I. Chugunov and D. G. Yakovlev, Astronomy Reports {\bf 49} (2005) 724.
\bibitem{viscospasta} C. J. horowitz and D. K. Berry, PRC {\bf 78} (2008) 035806.
\bibitem{Bdecay} P. Goldreich and A. Reisenegger, ApJ. {\bf 395} (1992) 250.








\end{thebibliography}
\end{document}